\documentclass[aps,prl,twocolumn,superscriptaddress,showpacs]{revtex4}

\usepackage{color}
\usepackage{graphicx}

\begin{document}

\input epsf.sty
\title{Novel in-gap spin state in Zn-doped La$_{1.85}$Sr$_{0.15}$CuO$_{4}$}


\author{H. Kimura}
\email[]{kimura@tagen.tohoku.ac.jp}
\affiliation{Institute of Multidisciplinary Research for Advanced Materials, 
Tohoku University, Sendai 980-8577, Japan}

\author{M. Kofu}

\author{Y. Matsumoto}

\author{K. Hirota}
\affiliation{Department of physics, Tohoku University, Sendai 980-8578, Japan}


\date{\today}

\begin{abstract}
Low-energy spin excitations of La$_{1.85}$Sr$_{0.15}$Cu$_{1-y}$Zn$_{y}$O$_{4}$ 
were studied by neutron scattering. In $y=0.004$, the incommensurate magnetic 
peaks show a well defined ``spin gap'' below $T_{\rm c}$.
The magnetic signals at $\omega=3\ {\rm~meV}$ decrease below 
$T_{\rm c}=27$~K for $y=0.008$, also suggesting the gap opening. At lower 
temperatures, however, the signal increases again, implying a novel 
{\em in-gap} spin state. In $y=0.017$, the spin gap vanishes and elastic 
magnetic peaks appear. These results clarify that doped Zn impurities induce the 
novel in-gap state, which becomes larger and more static with increasing Zn. 
\end{abstract}


\maketitle

It is widely accepted that the antiferromagnetism on a hole-doped CuO$_{2}$ 
plane in lamellar copper oxides is relevant to the high-$T_{\rm c}$ superconductivity. 
Therefore, a complete description of the interplay between the spin correlations and 
the dynamics of doped holes is indispensable to clarify the high-$T_{\rm c}$ mechanism.

The momentum and energy structure of antiferromagnetic (AF) spin correlations on 
the CuO$_{2}$ plane in La$_{2-x}$Sr$_{x}$CuO$_{4}$ (LSCO), which is a prototypical 
high-$T_{\rm c}$ superconductor, have been extensively studied by 
neutron scattering\cite{Kastner1998}. The spin excitations of the superconducting LSCO 
exhibit a quartet of peaks at the incommensurate wave vectors 
$Q_{\delta}=(\frac{1}{2}\pm\delta\ \frac{1}{2}\ 0), 
(\frac{1}{2}\ \frac{1}{2}\pm\delta\ 0)$ in the high temperature 
tetragonal (HTT) notation\cite{Cheong1991} and there exists 
a linear relation between $\delta$ and $T_{\rm c}$ in the 
underdoped region ($x\leq 0.15$)\cite{Yamada1998}. 
Neutron scattering studies have also revealed a well defined gap 
on spin excitation spectra, often called ``spin gap'', 
in LSCO\cite{Yamada1995,Lake1999,Lee2000} and 
YBa$_{2}$Cu$_{3}$O$_{7-\delta}$ (YBCO)\cite{Mignod1991,Sternlieb1993} 
around the optimally doped concentrations. Although the interrelations between the 
superconducting gap in the electronic state and the spin gap are not completely 
understood, the results of the neutron scattering studies indicate a 
strong relevance of the $q,\omega$-dependent spin excitations to the superconductivity 
and have contributed to the development of theoretical frameworks such as 
the {\em stripe} model\cite{Kivelson1998} and the {\em fermiology}\cite{Yamase2001}. 
However, the microscopic nature of spin correlations and their contributions to 
the high-$T_{\rm c}$ pairing mechanism still remain open questions. 

A small amount of doped Zn$^{2+}$ ions, resulting in the substitution for 
Cu$^{2+}$ ions, strongly suppress the superconductivity\cite{Xiao1990}. 
In addition, NMR studies\cite{Julien2000,Mahajan1994} have revealed that 
a Zn impurity induces staggered magnetic moments on Cu sites around the impurity, 
indicating that Zn-doping strengthens AF spin correlations on the CuO$_{2}$ plane. 
Neutron scattering studies have also revealed a drastic change 
in the low-energy spin dynamics: In Zn-free La$_{1.86}$Sr$_{0.14}$CuO$_{4}$, 
a gap-like nature in the spin excitations has been confirmed below 
$T_{\rm c}=33$~K\cite{Mason1993}, while the low-energy 
spin excitations still survive even below $T_{\rm c}=19$~K in La$_{1.86}$Sr$_{0.14}$Cu$_{0.988}$Zn$_{0.012}$O$_{4}$\cite{Matsuda1993}. 
Furthermore, the spin correlations become static while the incommensurate wave vector 
stays at $Q_{\delta}$\cite{Hirota1998}. Some recent theories concluded that the local antiferromagnetism is induced around non-magnetic impurities, where the superconductivity 
is locally suppressed\cite{Wang2002,Zhu2002}. 
These facts indicate the importance of microscopic coexistence and competition 
between the superconductivity and the AF order. 

A comprehensive neutron scattering study of the AF spin correlations 
in Zn-doped La$_{1.85}$Sr$_{0.15}$CuO$_{4}$ single crystals was performed to 
elucidate how the spin-gap state is broken 
and how the static AF correlations are induced 
by Zn-doping. To obtain quantitative information about the Zn-doping 
dependence of spin excitation spectra, it is essential to control the Zn-doping 
rate accurately. Furthermore, large and spatially homogeneous crystals are 
required because of the weak magnetic signals. 
We have overcome such difficulties by combining an improved 
traveling-solvent-floating-zone (TSFZ) method\cite{Lee1998} and 
a quantitative analysis of Zn impurities using the inductively-coupled plasma 
(ICP) method. The structural properties (size, shape, mosaicness, etc.) 
were also unified for all the samples so that the spin excitation spectra 
can be quantitatively compared among different samples. 
Systematic studies with changing the Zn-doping rate under unified experimental 
conditions revealed a novel low-energy spin excitation which is induced 
within the spin gap state by doped Zn impurities. 

Single crystals were grown by the traveling-solvent-floating-zone (TSFZ) method. 
The studied samples with a volume of  1~cm$^{3}$ were cut from the single crystal rods and 
properly annealed to eliminate oxygen deficiencies. 
\begin{table}[t]
\begin{center}
\setlength{\tabcolsep}{5pt}
\begin{tabular*}{3in}{ccccc}
 \hline
 Sample&\phantom{00}Sr $x$&Zn $y$&$T_{\rm c}$&$R_{\rm Zn-Zn}$\\ \hline
 $y=0.004$&\phantom{00}0.146(4)&0.004(1)&33$\pm 1$ K&60$\pm$7 \AA\\
 $y=0.008$&\phantom{00}0.147(4)&0.008(1)&28$\pm 1$ K&42$\pm$3 \AA\\
 $y=0.017$&\phantom{00}0.147(4)&0.017(1)&16$\pm 2$ K&29$\pm$1 \AA\\ \hline
\end{tabular*}
\end{center}
\caption{
Sr and Zn concentrations determined by ICP analysis and superconducting 
transition temperature $T_{\rm c}$ measured by the SQUID magnetometer. 
$R_{\rm Zn-Zn}$ denotes the mean distance between nearest-neighbor Zn atoms.
}
\label{table1}
\end{table}
The concentrations of Zn, Sr and Cu ions were precisely determined at several 
different points of each sample by a state-of-the-art ICP system (Shimadzu ICPS-7500), 
showing that Sr and Zn ions are doped homogeneously into the crystals. 
The obtained concentrations are listed in Table~\ref{table1}. 
$T_{\rm c}$ was determined from the shielding signal 
as a function of temperature using a SQUID magnetometer, which is in a good 
agreement with those of previous studies\cite{Xiao1990} for all the samples 
(See Table~\ref{table1}). The structural phase transition 
temperature $T_{\rm d1}$ from the high temperature tetragonal (HTT) to low 
temperature orthorhombic (LTO) phases was determined 
by neutron diffraction. Note that $T_{\rm d1}$ is quite sensitive 
to the Sr concentration. The obtained values are identical for all the samples 
($\simeq185$ K) and consistent with that of Zn-free LSCO of 
$x=0.15$\cite{Yamada1995}. The results indicate that the Sr concentration is 
exactly $x=0.15$ and that the Zn impurities do not affect the averaged crystal 
structure. 

Neutron scattering experiments were performed on the 
Tohoku University triple axis spectrometer (TOPAN) 
installed at JRR-3M in Japan Atomic Energy Research Institute (JAERI). 
The initial and final neutron energies were tuned by the Pyrolytic Graphite (PG) 
monochromator and fixed at 13.5 meV by the PG analyzer. A one-inch-thick 
PG filter was inserted in the scattered beam to eliminate higher-order contaminations. 
An additional PG filter was put in the incident beam for studying the elastic peaks. 
We mounted all the crystals in the ($h\ k\ 0$) zone and defined the 
reciprocal lattice unit (r.l.u.) in the HTT notation. 
In the present study, $q$-scans were performed around 
$\left(\frac{1}{2}\ \frac{1}{2}\ 0\right)$ at several different transfer 
energies $\omega$.  To normalize the data, we have utilized the acoustic 
phonons measured under the fixed condition because phonon intensity is 
considered proportional to the effective volume of a sample.

\begin{figure}
\centerline{\epsfxsize=2.6in\epsfbox{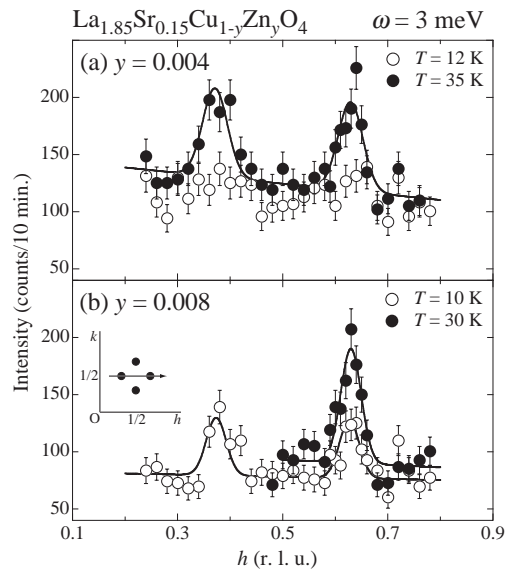}}
\caption{
$q$-profiles along the $h$-direction through ($\pi$, $\pi$) 
at $\omega=3$~meV for (a) $y=0.004$ and (b) $y=0.008$ 
below (Open circles) and above (Closed circles) $T_{\rm c}$.
}
\label{fig1}
\end{figure}
\begin{figure}
\centerline{\epsfxsize=2.6in\epsfbox{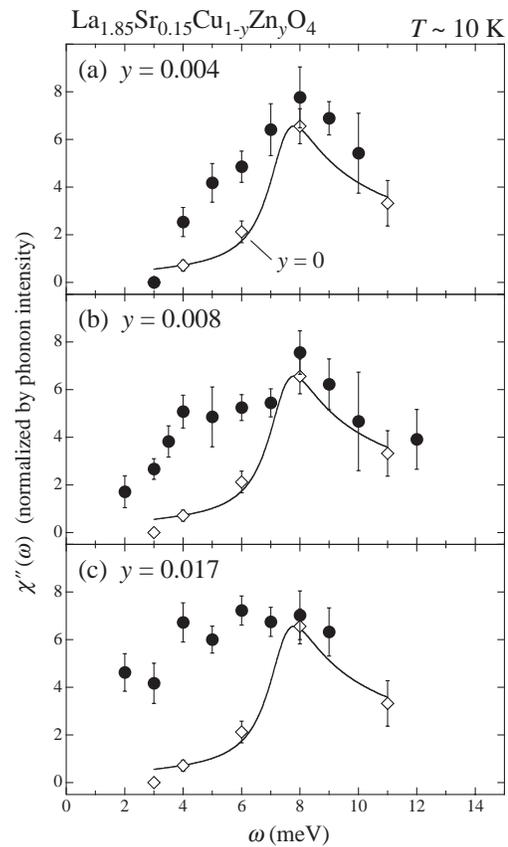}}
\caption{
Energy dependence of the $q$-integrated 
$\chi^{\prime\prime}(\omega)$ for (a) $y=0.004$, (b) $y=0.008$, and 
(c) $y=0.017$ taken around 10 K. Open diamonds in all the figure denote 
the data for $y=0$ taken in the present study. Solid lines are fits 
with a phenomenological dynamical spin susceptibility introduced in Ref.~\cite{Lee2000}.
}
\label{fig2}
\end{figure}
Figures~\ref{fig1}(a) and (b) show $q$-scan profiles at $\omega=3$~meV 
for $y=0.004$ and $y=0.008$, taken at 10~--~12~K (open circles) and 
just above $T_{\rm c}$ (closed circles). 
The trajectory of the scan is depicted in the inset of Fig.~\ref{fig1}(b). 
Above $T_{\rm c}$, the spin excitations have peaks at $Q_{\delta}$ 
with $\delta\sim0.12$ for both the samples. At 12~K, which is well 
below $T_{\rm c}$, the signal vanishes for $y=0.004$, implying the opening 
of the spin gap, while the intensity still remains at $Q_{\delta}$ for $y=0.008$. 
Energy spectra of the $q$-integrated dynamical spin susceptibility 
$\chi^{\prime\prime}(\omega)$ around 10~K are plotted in 
Fig.~\ref{fig2}(a)--(c) for $y=0.004$, 0.008, and 0.017. 
The open diamonds and the solid lines in the figures denote 
$\chi^{\prime\prime}(\omega)$ for Zn-free LSCO at $x=0.15$.
All the data are corrected with the thermal population factor and 
normalized by the intensity of acoustic phonon so that we can directly compare the 
amplitudes of $\chi^{\prime\prime}(\omega)$ for all the samples. 
It is remarkable that $\chi^{\prime\prime}(\omega)$ for all the samples 
have a maximum and almost identical intensity around $\omega=8$~meV while 
$\chi^{\prime\prime}(\omega)$ below 8~meV developes with increasing 
doped Zn. These systematic changes cannot be explained by either the 
broadening of the gap structure or the reduction of the gap-energy because both 
the cases should be associated with the variation of 
$\chi^{\prime\prime}(\omega)$ near the gap-energy. For example, 
in the case of the gap-broadening, 
the spectral weight just below the gap energy should increase 
while the weight above the gap energy decreases. Therefore, these energy 
spectra indicate that Zn-doping induces an {\em additional} spin excitation 
in the spin gap below $\omega \sim 8$~meV and that, with futher Zn-doping, 
the novel spin excitation is enhanced and shifts to lower energies, i.e., 
becomes more static. 

The additional spin excitations are also seen in the 
\begin{figure}[b]
\centerline{\epsfxsize=2.6in\epsfbox{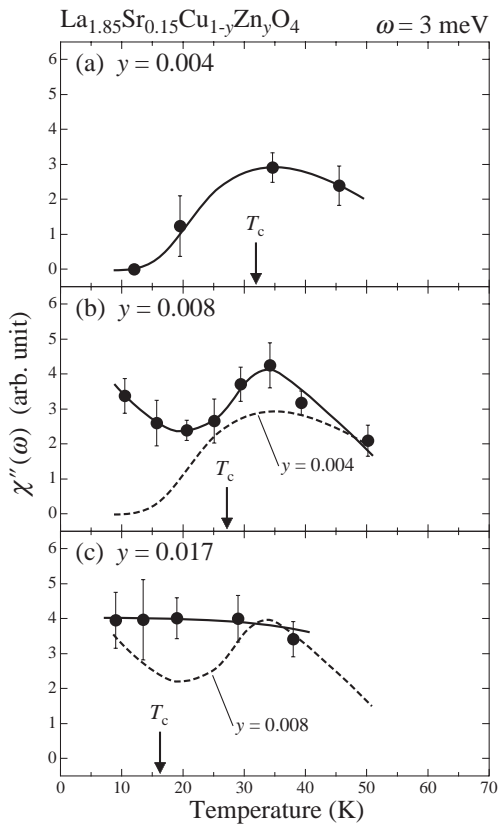}}
\caption{
Temperature dependence of the $q$-integrated 
$\chi^{\prime\prime}(3\ {\rm meV})$ for (a) $y=0.004$, (b) $y=0.008$ 
and (c) $y=0.017$. Solid lines in all the figures are guides to the eye.
}
\label{fig3}
\end{figure}
temperature dependence of the $\chi^{\prime\prime}(\omega)$ at 
$\omega=3$ meV ($\chi^{\prime\prime}(3\ {\rm meV})$). 
The results are summarized in Figs.~\ref{fig3}(a)-(c), showing a systematic 
variation of the spin excitations as a function of 
Zn-doping. In $y=0.004$, $\chi^{\prime\prime}(3\ {\rm meV})$ starts 
decreasing below $T_{\rm c}$ and goes to zero around 10 K, 
corresponding to the evolution of spin gap state. 
The $\chi^{\prime\prime}(3\ {\rm meV})$ of $y=0.008$ 
exhibits an interesting temperature dependence: As temperature is reduced, the $\chi^{\prime\prime}(3\ {\rm meV})$ once 
decreases around $T_{\rm c}$, which suggests the gap opening, 
but then increases {\em again} below $\sim20$~K. 
The low-temperature upturn indicates that the low-energy excitations 
by Zn-doping is a {\em novel} in-gap state, 
not simply due to a reduction of the gap energy. As shown in Fig.~\ref{fig3}(c), 
$\chi^{\prime\prime}(3\ {\rm meV})$ for $y=0.017$ is almost 
temperature independent around $T_{c}$, 
which is qualitatively consistent with the result of 
La$_{1.86}$Sr$_{0.14}$Cu$_{0.988}$Zn$_{0.012}$O$_{4}$\cite{Matsuda1993}, 
and suggests a complete vanishing of the spin-gap state. 

\begin{figure}
\centerline{\epsfxsize=2.55in\epsfbox{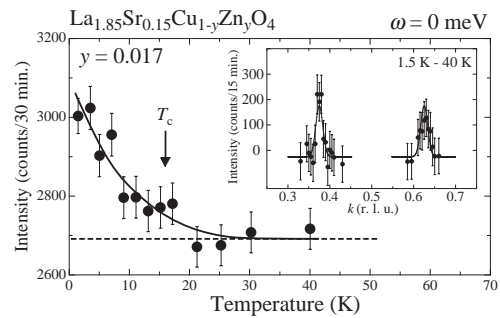}}
\caption{
Temperature dependence of the incommensurate elastic peak intensity of $x=0.017$. 
Solid line is guide to the eye. The inset shows the difference profile of 
the elastic magnetic peaks at 1.5 K and 40 K.
}
\label{fig4}
\end{figure}

Elastic scattering experiments were performed for $y=0.008$ and $y=0.017$ to 
investigate static spin correlations. In $y=0.008$, no signal 
was detected down to $T=1.5$~K, while in $y=0.017$, sharp elastic peaks were 
observed below $\sim 20$~K at the same incommensurate wave vector 
$Q_{\delta}$ as observed in the inelastic scattering measurements. 
Figure~\ref{fig4} shows temperature dependence of the elastic peak intensity 
for $y=0.017$. The inset shows the $q$-profile at 1.5~K with the 40~K data 
subtracted as background. An in-plane spin correlation length is estimated 
at $\sim 80$~\AA, which was obtained from the intrinsic line width of the 
$y=0.017$ peak profile. The results for $y=0.008$ and $y=0.017$ suggest 
that the novel spin state in $y=0.008$ is 
purely dynamical and becomes more static with increasing Zn-doping. 

The present study has shown that the reduction of $T_{\rm c}$ and the 
development of antiferromagnetic correlations are continuously tunable 
by successively doping Zn impurities, 
where the superconductivity and AF ground state 
competitively coexist with each other. 
Energy dependence of $\chi^{\prime\prime}(\omega)$ shows that 
Zn-doping enhances a low-energy spin excitation while no significant 
variation occurs around the $\omega=8$~meV region where the gap 
starts opening in Zn-free LSCO. Furthermore, the temperature dependence 
of $\chi^{\prime\prime}(3\ {\rm meV})$ for $y=0.008$ indicates that 
with decreasing temperature, the induced spin excitations are 
followed by an opening of the spin gap. These two facts imply that 
a Zn-doping yields a novel in-gap spin state instead of the broadening of the 
gap-edge or the reduction of the gap-energy. 
In the YBCO system, Zn-doping also 
induces a low-energy spin excitation which coexists 
with a gap-like feature, suggesting two kinds of copper sites; one around 
Zn ions and the other almost Zn-independent\cite{Sidis1996}. 
This result is consistent with that of NMR studies\cite{Julien2000,Mahajan1994}, 
showing that the local moments are induced at the copper sites around Zn ions. 
In $y=0.008$, a mean distance between Zn ions 
($\equiv R_{\rm Zn-Zn}$) shortened from $\sim 60$ \AA\ in $y=0.004$ to 
$\sim 42$ \AA\ (See Table~\ref{table1}). Thus we speculate that 
the induced local magnetic moments around a doped Zn ion start correlating 
with those around other Zn ions for $y=0.008$, and that the correlations among 
the moments around different Zn ions become coherent, which gives rise to the 
novel in-gap spin state at particular $q$ positions, i.e., the 
$q$-dependent spin excitations near the zero energy. 
In $\mu$SR studies, Nachumi {\it et al}.\cite{Nachumi1996} proposed 
a ``swiss cheese'' model in which charge carriers in an area of $\pi\xi_{ab}^{2}$ 
($\xi_{ab}\sim 18$ \AA) around Zn impurities are excluded from 
the superconductivity. These results support a picture that the superconductivity 
is locally destroyed by the induced moments around Zn impurities but still survives. 
This might give a possible explanation for the microscopic coexistence of the 
superconductivity and the antiferromagnetism, in the form of an inhomogeneous 
mixture of these two ground states\cite{Wang2002,Zhu2002}. 

Static spin correlations characterized by the incommensurate elastic magnetic peaks 
are observed in $y=0.014$\cite{Hirota1998_2} and 0.017, where the $R_{\rm Zn-Zn}$ 
values are 32 \AA\ and 29 \AA, respectively. The in-plane spin correlation lengths for 
$y=0.014$ and 0.017 exceed 80 \AA\ which is much longer than those of $R_{\rm Zn-Zn}$ 
and $\xi_{ab}$ in the $\mu$SR study\cite{Nachumi1996}. 
These facts show that the static correlations originate {\em not} from the independent 
local magnetisms around Zn-impurities {\em but} from the long-range AF coherence among 
the induced moments around different Zn ions. In addition, the elastic magnetic peaks for 
$y=0.014$ and $y=0.017$ have the same incommensurate wave vector $Q_{\delta}$ as 
that of the in-gap spin excitations in $y=0.008$. 
Thus we conclude that the in-gap state continuously connects to an AF ground 
state with increasing Zn impurities, i.e., with decreasing $R_{\rm Zn-Zn}$. 
We note that these results are contrast to the case of LSCO at $x=0.12$, where 
3~\% of Zn-doping not only completely suppresses the superconductivity but also 
disturbs the long-range AF order\cite{Kimura1999}. However, recent $\mu$SR studies 
for LSCO of $x=0.115$, which has static spin correlations, showed that with 
increasing Zn, the spin correlations are primarily enhanced but destroyed 
by further Zn-doping\cite{Watanabe2002}. In the present case, Zn-free LSCO ($x=0.15$) 
shows the spin gap state, indicating that there is no long-range order as a ground state. 
Therefore further doping of Zn is required for stabilizing a long range order 
than that required for $x\sim0.12$. 

We finally quote the field-induced low-energy 
spin excitations on La$_{1.837}$Sr$_{0.163}$CuO$_{4}$ found by 
Lake {\it et al}.\cite{Lake2001}. They argued that the induced excitations 
originate from the network among vortex cores in which the spin correlations are 
antiferromagnetic. These results are relevant to our results and imply that the 
superconductivity can competitively coexist with an AF ground state 
which is introduced by the local impurities or vortices. 

In conclusion, we have studied low-energy spin correlations in the systematically 
Zn-doped La$_{1.85}$Sr$_{0.15}$CuO$_{4}$ samples. 
We found in $y=0.008$ that a novel in-gap spin state around 3 meV 
develops on cooling temperature, which corresponds to 
the intermediate state between the spin gap and the AF order. 
The systematic variation from the reduction of the spin-gap state to the 
emergence of the static spin correlations via the novel in-gap spin state 
is consistent with a competitively coexistence in the form of an 
inhomogeneous mixture of superconducting regions and AF 
regions. The present study shows the importance of underlying AF ground state which is 
locally substituted for the superconducting state with help of small perturbations. 

We thank Y. Endoh, M. Fujita, H. Hiraka, C. H. Lee, M. Matsuda, G. Shirane, 
J. M. Tranquada, H. Yamase, and K. Yamada, for stimulating discussions. 
This work was supported in part by aGrant-In-Aid for Encouragement of 
Young Scientists (13740198, 2001) from the Japanese Ministry of
Education, Science, Sports and Culture, and by the Core Research for Evolutional Science 
and Technology (CREST) from  the Japan Science and Technology Corporation. 
\begin{acknowledgments}
\end{acknowledgments}


\begin{references}

\bibitem{Kastner1998}M. A. Kastner {\it et al}., 
Rev. Mod. Phys. {\bf 70}, 897 (1998), and references therein.

\bibitem{Cheong1991}S. -W. Cheong {\it et al}., 
Phys. Rev. Lett. {\bf 67}, 1791 (1991).

\bibitem{Yamada1998}K. Yamada {\it et al}., 
Phys. Rev. B {\bf 57}, 6165 (1998).

\bibitem{Yamada1995}K. Yamada {\it et al}., 
Phys. Rev. Lett. {\bf 75}, 1626 (1995).

\bibitem{Lake1999}B. Lake {\it et al}., 
Nature {\bf 400}, 43 (1999).

\bibitem{Lee2000}C. H. Lee {\it et al}., 
J. Phys. Soc. Jpn. {\bf 69}, 1170, (2000).

\bibitem{Mignod1991}J. Rossat-Mignod {\it et al}., 
Physica C {\bf 185-189}, 86 (1991).

\bibitem{Sternlieb1993}B. J. Sternlieb {\it et al}., 
Phys. Rev. B {\bf 47}, 5320 (1993).

\bibitem{Kivelson1998}S. A. Kivelson {\it et al}., 
Nature {\bf 393}, 550 (1998).

\bibitem{Yamase2001}H. Yamase {\it et al}., 
J. Phys. Soc. Jpn. {\bf 70}, 2733, (2001).

\bibitem{Xiao1990}G. Xiao {\it et al}., 
Phys. Rev. B {\bf 42}, 8752 (1990).

\bibitem{Julien2000}M. -H. Julien {\it et al}., 
Phys. Rev. Lett. {\bf 84}, 3422 (2000).

\bibitem{Mahajan1994}A. V. Mahajan {\it et al}., 
Phys. Rev. Lett. {\bf 72}, 3100 (1994).

\bibitem{Mason1993}T. E. Mason {\it et al}., 
Phys. Rev. Lett. {\bf 71}, 919 (1993).

\bibitem{Matsuda1993}M. Matsuda {\it et al}., 
J. Phys. Soc. Jpn. {\bf 62}, 443, (1993).

\bibitem{Hirota1998}K. Hirota {\it et al}., 
Physica B {\bf 241-243}, 817 (1998).

\bibitem{Wang2002}Z. Wang {\it et al}., 
Phys. Rev. Lett. {\bf 89}, 217002 (2002).

\bibitem{Zhu2002}J. -X. Zhu {\it et al}., 
Phys. Rev. Lett. {\bf 89}, 067003 (2002).

\bibitem{Lee1998}C. H. Lee {\it et al}., 
Supercond. Sci. Technol. {\bf 11}, 891 (1998).

\bibitem{Sidis1996}Y. Sidis {\it et al}., 
Phys. Rev. B {\bf 53}, 6811 (1996).

\bibitem{Nachumi1996}B. Nachumi {\it et al}., 
Phys. Rev. Lett. {\bf 77}, 5421 (1996).

\bibitem{Hirota1998_2} The Zn concentration of the crystal studied in 
Ref.~\cite{Hirota1998} was originally assigned to $y=0.012$. 
We reanalyzed this crystal with ICP and found $y=0.014$.

\bibitem{Kimura1999}H. Kimura {\it et al}., 
Phys. Rev. B {\bf 59}, 6517 (1999).

\bibitem{Watanabe2002}I. Watanabe {\it et al}., 
Phys. Rev. B {\bf 65}, 180516 (2002).

\bibitem{Lake2001}B. Lake {\it et al}., 
Science {\bf 291}, 1759 (2001).

\end{references}

\end{document}